\documentclass{iau} 
\usepackage{graphicx}
\newcommand\ocen{$\omega$\,Cen}

\newcommand\msun{\rm{M$_{\odot}$}}
\newcommand\teff{T$_{\rm eff}$}  
\newcommand\Teff{T$_{\rm eff}$}  

\title[The AGB scenario for multiple populations] 
{`Status update' for the AGB scenario for the formation of multiple populations}

\author[Francesca D'Antona]   
{Francesca D'Antona}

\affiliation{INAF-OAR\\
I-00078 Monte Porzio Catone, Italy \\ email: {\tt francesca.dantona@inaf.it} 
}

\pubyear{2019}
\volume{351}  
\jname{Star Clusters: From the Milky Way to the Early Universe}
\editors{A. Bragaglia, M.B. Davies, A. Sills \& E. Vesperini, eds.}
\begin{document}

\maketitle

\begin{abstract}
The Asymptotic Giant Branch (AGB) scenario for the formation of multiple populations has been quantitatively studied in the course of the latest twenty years, examining the aspects both of nucleosynthesis and of the dynamics of formation of new stars in a cooling flow at the center of the first generation cluster, and of the following N--body evolution.  The large complex of these studies finds many validations in the properties of multiple populations. 
Here I shortly summarize recent accomplishments in the study of the evolution of massive AGBs and super--AGBs 
including the the explanation of anomalous high lithium abundances in `extreme' second generation stars in \ocen\ and NGC\,2808. 
\keywords{globular clusters: general, stars: AGB and post-AGB}
\end{abstract}

\firstsection 
\section{Introduction}

More than ten year ago the  AGB scenario for multiple populations (MPops) was presented in the seminal paper by \cite[D'Ercole\,et\,al.\,2008]{dercole2008},  including a 1D dynamical model, N--body simulations and chemical predictions concerning helium. The cooling flow determined by the collection in the cluster core of the low--speed ejecta of the most massive AGB stars, diluted with re-accreted pristine matter, formed second--generation (2G) stars whose chemistry grossly reproduced the patterns of abundances of Globular Clusters (GC) stars. In 2008, the idea that AGB ejecta could contribute to the anomalies in GCs was already more than 25 years old (\cite{cottrell-dacosta1981, dantona1983}), but it had been successfully revitalised, when the chemical anomalies were observed in scarcely evolved GC stars (e.g. \cite{gratton2001, cohen2002}), thanks to the contemporary  unexpected results of the nucleosynthesis in Hot Bottom Burning (HBB) of massive AGBs of low metallicity (\cite[Ventura et al.\,2001]{ventura2001}). The high temperatures reached at the bottom of the convective envelope in these models showed that the oxygen envelope was depleted by the action of the ON cycle, a feature which had not been found or highlighted as significant in previous computations.\\
We must be aware that, in the same way as ladies' outfits are valuable if they include `a decent dress', and not only beautiful shoes or an appealing hat, the main ingredient (the dress) of an MPops model are the chemical patterns it provides. The AGB scenario has been developed from two complementary points of view: 
\begin{itemize}
\item{\sl the fabric of the dress}, that is the exploration of the chemical patterns of the ejecta from full stellar models of AGBs, to be compared with the GC patterns. The dependence of yields both on the evolving mass and on the metallicity is a further property of the model;
\item{\sl the design of the dress}, that is the dynamical formation of  the 2G  in a cooling flow, the following N--body evolution of the 1G--2G spatial distributions and the properties of different populations (e.g. the binaries)
\end{itemize} 
In spite of its relevant and extensive successes, in part documented in this Symposium, anyway, the most popular advertisement today is that ``all models so far proposed fail to explain all the characteristics of multiple populations" posing the AGB scenario at the same level of models excluded by patent inability to comply with the main properties of MPops, and in particular, with the chemical patterns. 

\section{Massive AGB and super--AGB evolution: warning on the models and on the narration of model results}
One problem which apparently undermines the AGB scenario is that the AGB model building has to deal with several problematic aspects of micro-- and macro--physics. \\
Concerning the micro--physics, an important aspect is the role of nuclear reaction rates, whose difficulties have been dealt with in several works (e.g. \cite{ventura2006}, \cite[Renzini et al.\,2015]{renzini2015}, \cite{dantona2016}, \cite{ventura2018}). An open problem is to ascertain better, in future experiments,  some important reaction rates. In some cases, the actual uncertainty may be larger than tabulated. As an example, for the $^{23}$Na(p,$\alpha$)$^{20}$Ne reaction, which is critical to determine the abundance of sodium in the ejecta, the uncertainties on the reaction rate in the astrophysical range 70--110\,MK may have been underestimated due to the presence of a not yet well studied resonance at 138\,Kev.\\
For the macro--physics, both mass loss law and the convection model, whose formulations do not come from first principles, are highly uncertain in these phases, also due to the lack of observational guidelines. \\
Successful models for the 2G require that the HBB phase is not too long lived, to avoid dramatic effects from the 3rd dredge-up (\cite{ventura2005a}), but long enough that it allows ON cycling and Mg processing (e.g.\,\cite{ventura2018}). \\
Convection must be very efficient to produce an effective HBB.
Scarcely efficient convection ---such as modelled by assuming in the AGB a solar-calibrated Mixing Length Theory (MLT) ratio of mixing length to pressure scale height--- produces too low HBB temperatures and a low degree of p--capture processing. In addition, the total luminosity for a given stellar mass (in the range of possible polluter masses) is lower for lower convection efficiency, so the evolution is longer and there is a stronger effect of the 3rd dredge up (\cite{ventura2005b}), all features which lead far away from consistency with observed chemical anomalies.\\
Due to the sensitivity of yields on the mass loss, and especially on the convection modelling, one typical criticism of the AGB scenario is that ``yields of AGBs differ amongst different authors" and so reach, or not, the ability to reproduce the observed light elements patterns (see, e.g., Charbonnel 2016), but let me shift this reasoning (models differ according to who computes them) to the computation of the {\sl solar model}. All researchers, and also the students of a basic stellar structure course, are aware that only a {\sl precise calibration} of convection allows to get the correct solar \Teff\ at the solar age, but nobody would feel the duty to say that ``the tracks of 1\msun\ differ amongst different authors'', simply because authors a priori calibrate convection on their own model to fit precisely the solar \teff.  (And in spite of the accurate calibration of the standard MLT solar model the envelope temperature stratification does not reproduce the high frequency solar oscillations patterns. In fact, the MLT calibration provides an algorithm to compute the ``average" temperature gradient along the whole convective region). \\
Concerning the AGB models, we should paradoxically conclude that, in the end, {\sl we could use GC abundance anomalies to ``calibrate" the convection model and mass loss laws for massive, low metallicity AGBs}.

\section{The timeline of formation of multiple populations}   
\begin{figure}[t ]
\vspace*{-3.2 cm}
\begin{center}
\includegraphics[width=3.2in]{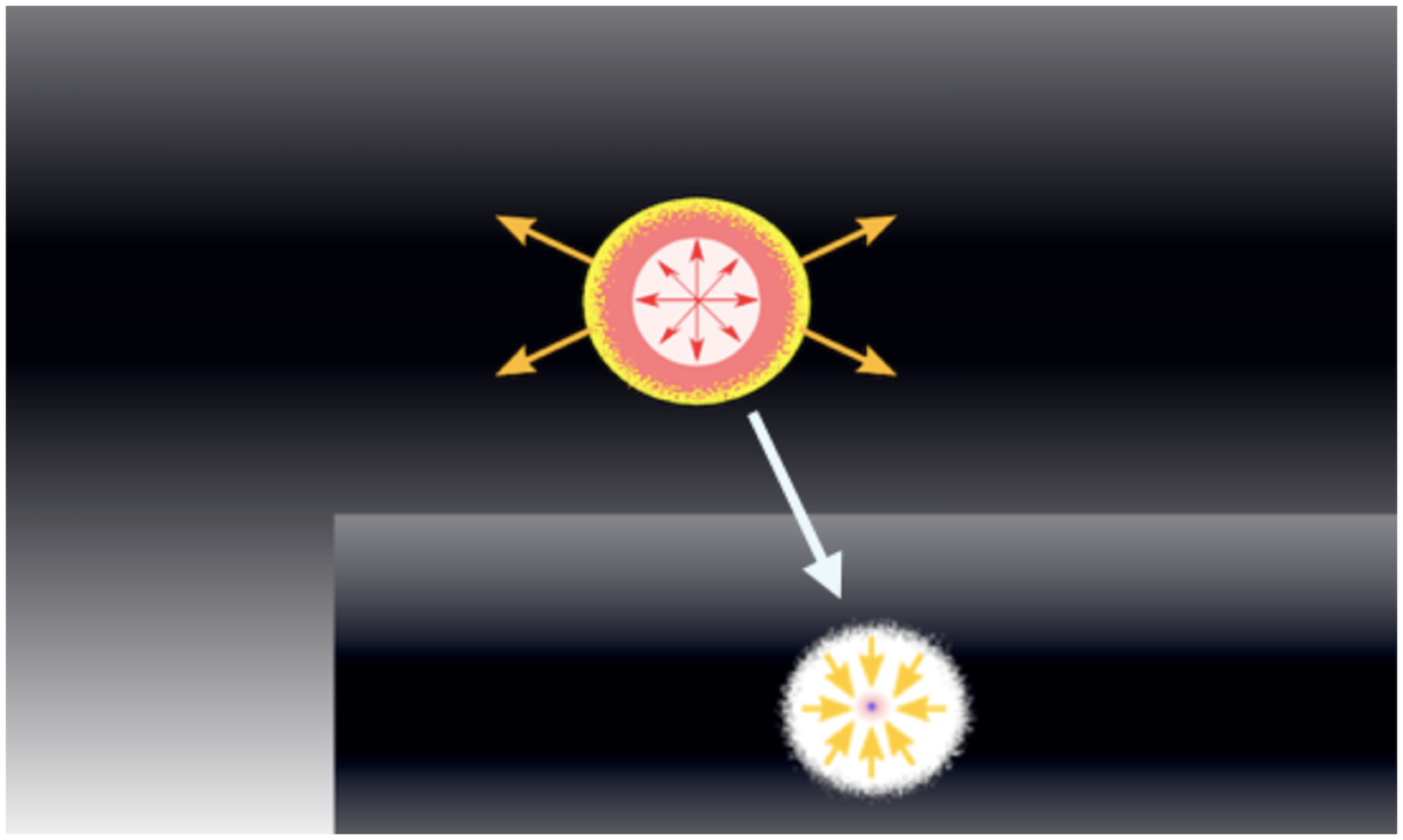} 
\vspace*{-3.5 cm}
\caption{Schematic effect of delayed type II supernovae on the star formation timeline. The bubble triggered by delayed SN II events is not able to expand out of the dwarf galaxy disk, and  SN and  AGB ejecta collect at the bubble limit (upper sketch). At the end of the delayed SN\,II epoch, the bubble closes (lower sketch), and star formation takes place in gas polluted by both AGB and SN ejecta. Type II clusters, showing spreads in iron, increased s--process elements and C+N+O  may form through this path (D'Antona et al. 2016, D'Ercole et al. 2016). }
\label{f1}
\end{center}
\end{figure}

Recently the classic AGB scenario by \cite[D'Ercole et al.\,(2008)]{dercole2008} was updated by \cite[D'Ercole et al.\,(2016)]{dercole2016} considering GC formation in the disk of a proto-dwarf galaxy, so dynamically solving both the problem of full loss of the pristine gas during the Supernova epoch and the problem of later re-accretion of pristine gas. In the same context, \cite[D'Antona et al.\,(2016)]{dantona2016} showed that  the variety and discreteness of GC MPops can be a byproduct of this scenario, with cluster-to-cluster difference arising from the precise timeline along which the MPops are formed. This model in particular presents a solution for the Type\,II clusters, by including the role of delayed type II supernovae (the SN events taking place by binary evolution after the end of the major SN epoch which cleared the cluster from the gas residual from the formation of the first generation). The explosion of these delayed SN may be able to forbid star formation in the cluster core, but not to push the re-accreting gas our of the cluster (see Figure\,\ref{f1}). When the delayed SN epoch ends, there will be a burst of star formation in matter enriched both in iron by the SN ejecta, and in CNO and s-process elements by the winds of the less massive AGBs (~4--4.5\msun) evolved during this epoch.

\section{Status update of models: Magnesium and Lithium}
\begin{figure}[t]
\vspace*{-3.2 cm}
\begin{center}
\includegraphics[width=3.8in]{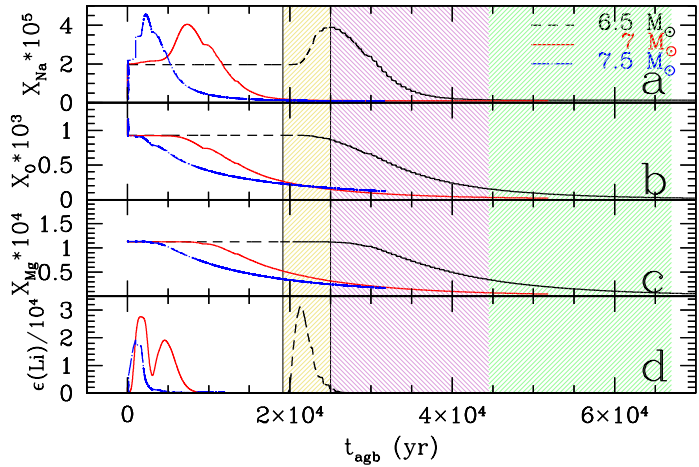} 
\vspace*{-0.2 cm}
 \caption{Surface abundances of Li, O, Mg and Na along the evolution of super--AGBs of masses 6.5, 7 and 7.5\msun, from \cite[D'Antona et al.\,2019]{dntona2019} }
   \label{f2}
\end{center}
\end{figure}

Recent sets of computations have made serious attempts to comply with open problems in the predictions of AGB nucleosynthesis. In particular, \cite{dicrisci2018} showed that better agreement with the Mg depletion in the extreme stars of NGC\,2808 can be achieved if during the super--AGB evolution the mass loss is modulated by the effects of radiation pressure on the dust grains.
The lithium yields from these same models are consistent with the high lithium abundance in a star (\cite[D'Orazi et al. 2015]{dorazi2015}), belonging to the `extreme' group of NGC\,2808 (\cite[D'Antona et al. 2019]{dantona2019}). Figure\,\ref{f2} shows that lithium is produced by the \cite{cameronfowler1971} chain during the first phases of super--AGB evolution. Remember that, in the AGB scenario, the extreme stars (present in NGC\,2808 and in a few other GCs, such as \ocen) are identified with stars born directly from the pure super--AGB ejecta. Notice that the epoch of high lithium in the envelope (panel d) is concomitant with the epoch of maximum sodium, corresponding to the HBB of the $^{22}$Ne brought in the envelope at the 2nd dredge up. Therefore, {\sl if there are extreme stars born from pure super--AGB ejecta}, and no additional mixing mechanisms have depleted in part their lithium, we may expect to find a direct correlation Na--Li. In fact a direct Na--Li correlation has been found in the two extreme stars examined by \cite[Mucciarelli et al. (2019)]{mucciarelli2019} in $\omega$Cen, whose extreme Lithium abundances are compatible only with the enormous Lithium production displayed by the models in \cite[D'Antona et al. (2012)]{dantona2012li} for super--AGBs close to the maximum mass not igniting as supernova. 
This peculiar behaviour of a few high lithium `extreme' stars should not be confused with the more common rule of an anticorrelation Na--Li present in 2G stars, which is due to the dilution of Na--rich ejecta with the Na--poor, but Li--rich intracluster pristine gas.  

\end{document}